\documentclass[aps,prd,twocolumn,
preprintnumbers,amsmath,amssymb,
epsf,superscriptaddress,groupedaddress,tightenlines,nofootinbib]{revtex4}
\usepackage{epsfig,graphicx}

\usepackage{graphicx}
\usepackage{amsmath}
\usepackage{bm}

\newcommand{\beqa}{\begin{eqnarray}}
\newcommand{\eeqa}{\end{eqnarray}}
\newcommand{\beq}{\begin{equation}}
\newcommand{\eeq}{\end{equation}}
\newcommand{\bsp}{\begin{split}}
\newcommand{\esp}{\end{split}}
\newcommand{\bal}{\begin{align}}
\newcommand{\eal}{\end{align}}
\renewcommand{\Re}{{\cal R}e}

\begin{document}

\preprint{SLAC-PUB-12063}
\preprint{BNL-HET-06/8}


\vspace*{10pt}
\title{Improved method for CKM constraints in charmless three-body $B$ and $B_s$ decays
\footnote{Research supported in part by the US Department of Energy, contract
DE-AC02-76SF00515}}
\def\addbnl{Physics Department, Brookhaven National Laboratory, Upton, New York 11973}
\def\addcmu{
Department of Physics, Carnegie Mellon University,
Pittsburgh, PA 15213}
\def\addIJS{J.~Stefan Institute, Jamova 39, P.O. Box 3000, 1001
Ljubljana, Slovenia}
\def\addtech{Physics Department, Technion--Israel Institute of Technology
32000 Haifa, Israel}
\def\addslac{Stanford Linear Accelerator Center, Stanford University, Stanford, CA 94309}
\def\addmit{Center for Theoretical Physics, Massachusetts Institute for Technology,
Cambridge, MA 02139}

\author{Michael Gronau}\affiliation{\addtech}\affiliation{\addslac}
\author{Dan Pirjol}\affiliation{\addmit}
\author{Amarjit Soni}\affiliation{\addbnl}
\author{Jure Zupan} \affiliation{\addcmu}\affiliation{\addIJS}

\begin{abstract} \vspace*{18pt}
Recently Ciuchini, Pierini and Silvestrini proposed a method for
constraining CKM parameters in $B\to K\pi\pi$ and $B_s\to
K\pi\pi$ through phase measurements of amplitudes involving $I=3/2$
$K^*\pi$ final states. We show that complementary information on CKM
parameters may be obtained by studying the phases of $\Delta I=1$ $B\to
(K^*\pi)_{I=1/2}$, $B_s\to (K^*\bar K)_{I=1}$ and $B_s \to (\bar
K^* K)_{I=1}$ amplitudes. Hadronic uncertainties in these constraints
from electroweak penguin operators $O_9$ and $O_{10}$, studied using flavor
SU(3), are shown to be very small in $B\to K\pi\pi$ and $B_s\to K\pi\pi$ and somewhat
larger in $B_s\to K\bar K\pi$. The first processes imply a precise linear relation
between $\bar\rho$ and $\bar\eta$, with a measurable slope and
an intercept at $\bar\eta=0$
involving a theoretical error of
0.03. The decays $B_s\to K\pi\pi$ permit a measurement
of $\gamma$ involving a theoretical error
below a degree.
We note that while time-dependence is required
when studying $B^0$ decays at the $\Upsilon(4S)$, it may not be needed when studying
$B_s$ decays at hadronic colliders.
\end{abstract}
\maketitle
\section{Introduction}
Recently a method has been proposed by Ciuchini, Pierini and Silvestrini 
\cite{Ciuchini:2006kv,Ciuchini:2006st} for
determining Cabibbo-Kobayashi-Maskawa (CKM) parameters in
three body $B\to K\pi\pi$ and $B_s\to K\pi\pi$ decays.
The proposed method is reminiscent of early suggestions for
determining $\gamma$ using rates and asymmetries in two body decays
$B\to K\pi$~\cite{Nir:1991cu,Gronau:1994bn,Fleischer:1997um,Neubert:1998pt}
and $B_s\to K\pi$~\cite{Gronau:2000md}.
A unique feature of the new method is being able to measure through
interference in the Dalitz plot relative phases between quasi
two-body decay amplitudes for $B_{(s)}\to K^*\pi$ and $\bar
B_{(s)}\to \bar K^*\pi$.
This is similar to a proposal for measuring relative phases
among $B\to \rho\pi$ amplitudes by studying the Dalitz plot for
$B^0\to\pi^+\pi^-\pi^0$~\cite{Snyder:1993mx}.
When neglecting electroweak penguin (EWP)
contributions, the relative phase between a combination of decay
amplitudes describing $B_{(s)}\to (K^*\pi)_{I=3/2}$ and a
corresponding combination of $\bar B_{(s)}$ amplitudes
determines the weak phase $\gamma$. A small hadronic uncertainty
caused by EWP amplitudes was estimated, based on factorization and
assuming certain input values for $B$-to-light-mesons form
factors~\cite{Ciuchini:2006kv}.

In the present paper we propose extending
the method to $\Delta I=1$, $I(K^*\pi)=1/2$ amplitudes in the
above decays and to $I=1$ amplitudes in
$B_s\to K^*\bar K$ and $B_s\to \bar K^* K$.
We use flavor SU(3) to study theoretical
uncertainties caused by EWP contributions,
suggesting a way for reducing these uncertainties.
The resulting theoretical precision in determining CKM parameters 
in $B\to K\pi\pi$ and $B_s\to K\pi\pi$ is shown to be very high, essentially at 
the level of isospin breaking corrections. This happens because the method 
is based primarily on isospin symmetry considerations, while flavor SU(3) is 
used only to estimate uncertainties from a subset of small EWP contributions.
 
In Section II we analyze $B\to K^*\pi$, $B_s\to K^*\pi$ and
$B_s\to K^*\bar K (\bar K^* K)$ decays in terms of
isospin amplitudes, selecting several ratios of $\bar B_{(s)}$ and $B_{(s)}$
isospin amplitudes which can be used to determine $\gamma$ in the absence
of EWP contributions. Section III studies the effects of EWP amplitudes, turning
the determination of $\gamma$ into a generic constraint on CKM
parameters. The constraint involves an uncertainty from a ratio of two hadronic matrix
elements
of $(V-A)$ current-current operators. 
Flavor SU(3) calculations show that this ratio is  small for
judiciously chosen combinations of isospin amplitudes in $B\to K^*\pi$, vanishes
approximately in $B_s\to K^*\pi$ in the isospin symmetry limit, and is larger in
$B_s\to K^*\bar K (\bar K^*K)$. This  implies precise constraints on CKM parameters
from knowledge of amplitudes and their relative phases for $B\to K^*\pi$  and
$B_s\to K^*\pi$. 

Section IV discusses measurements of these quasi two-body decay
amplitudes and of $B\to K^*\bar K~(\bar K^*K$) in three classes of three body decays,
$B\to K\pi\pi$, $B_s\to K\pi\pi$ and $B_s \to K\bar K\pi$ decays, respectively.
We point out that measuring a relative phase between the amplitudes for 
$B^0\to K^{*+}\pi^-$ and $\bar B^0\to K^{*-}\pi^+$ in $B^0\to K_S\pi^+\pi^-$ produced 
in $e^+e^-$ collisions at the $\Upsilon(4S)$ requires time-dependence. In contrast,  no
time-dependence may be needed for a similar measurement in $B_s\to K_S\pi^+\pi^-$ 
performed at hadronic colliders if a width difference in the $B_s$ system is measured.
In order to obtain a most precise determination of CKM parameters, we propose
applying amplitude analyses to the entire $B\to K\pi\pi$ class, using isospin amplitudes
as variables.  Section V concludes with several remarks
about the implementation of this method and its sensitivity to physics beyond the Standard 
Model, comparing it with two other methods for determining $\gamma$ in $B$ and $B_s$ 
decays. 

\section{Isospin decompositions and $\gamma$ without electroweak
penguin terms}\label{sec:decomposition}
The cleanest method for extracting the weak phase $\alpha$ or
$\gamma$ in $\Delta S=0$ and $\Delta S=1$ charmless hadronic $B$
decays stems from applying isospin symmetry to these decays,
eliminating the effect of QCD penguin amplitudes which transform in these
processes as $\Delta I=1/2$ and $\Delta I=0$, respectively~\cite{Gronau:1990ka}.
We will now discuss separately the four cases, $B\to K^*\pi$, $B_s\to K^*\pi$,
$B_s \to K^* \bar K$ and $B_s\to \bar K^* K$ in terms of their isospin amplitudes.

\subsection{$B\to K^*\pi$}
In strangeness changing decays of the type $B\to K\pi$ or $B\to
K^*\pi$, where $K^*$ denotes any kaon resonance, $K^*(892), K^*_0(1430),
K_2^*(1430), \dots$, the four physical amplitudes for $B^0$ and $B^+$ are
decomposed into three isospin-invariant amplitudes~\cite{Nir:1991cu},
\begin{align}
\begin{split}
-A(K^{*+}\pi^-) &= B_{1/2} - A_{1/2} - A_{3/2}~,\\
A(K^{*0}\pi^+) &= B_{1/2} + A_{1/2} + A_{3/2}~,\\
-\sqrt{2}A(K^{*+}\pi^0) &= B_{1/2} + A_{1/2} - 2A_{3/2}~,\\
\sqrt{2}A(K^{*0}\pi^0) &=B_{1/2} - A_{1/2} + 2A_{3/2}~.
\end{split}
\end{align}
Here we use a phase convention~\cite{Gronau:1994rj} in which a minus sign
is associated with a $\bar u$ quark in a meson. The amplitudes $B$ and $A$
correspond to $\Delta I=0$ and
$\Delta I=1$ parts of ${\cal H}_{\rm eff}$, respectively. Their subscripts denote
the isospin of the final $K^*\pi$ state. Here and elsewhere we will denote by $B$
isospin amplitudes obtaining contributions from QCD penguin operators, and by $A$ other
amplitudes. Our study will focus on the latter.

The amplitude quadrangle relation,
\begin{align}
\begin{split}
3A_{3/2} &= A(K^{*+}\pi^-) +\sqrt{2}A(K^{*0}\pi^0)\\
&= A(K^{*0}\pi^+) + \sqrt{2}A(K^{*+}\pi^0)~,
\end{split}\label{quad}
\end{align}
defines $A_{3/2}$ as one diagonal of the quadrangle, while $A_{1/2}$ is
given by
\begin{align}
\begin{split}
6A_{1/2} &= A(K^{*+}\pi^-) + 3A(K^{*0}\pi^+)- 2\sqrt{2}A(K^{*0}\pi^0)\\
&= 3A(K^{*+}\pi^-) + A(K^{*0}\pi^+) - 2\sqrt{2}A(K^{*+}\pi^0)~.
\end{split}\label{A1/2}
\end{align}

The two $\Delta I=1$ amplitudes, $A_{3/2}$ and $A_{1/2}$, do not
contain a QCD penguin contribution and would carry a single weak phase
$\gamma$ if EWP contributions could be neglected. Here we will
proceed under this assumption, postponing a discussion of the
effects of EWP amplitudes to the next section. Denoting amplitudes
for charge-conjugate initial and final states by $\bar A$, and
defining two ratios of amplitudes,
\beq\label{RI}
R_I \equiv
\frac{\bar A_I}{A_I}~,~~~~I=1/2,~3/2~,
\eeq
the phase $\gamma$ is
determined by
\beq\label{argRI}
\Phi_I\equiv -\frac{1}{2}{\rm arg}(R_I) = \gamma~.
\eeq
Note that although the ratios $R_I$ do not depend on the magnitudes of
$A_I$ in the limit of vanishing EWP contributions, an extraction of $\gamma$
requires measuring both the magnitudes and the relative phases of physical
$B\to K^*\pi$ amplitudes and their charge-conjugate.

The ratio $R_{3/2}$ was studied in~\cite{Ciuchini:2006kv} (where it
was denoted by $R^0=R^{\pm}$) while the ratio $R_{1/2}$ studied here
provides independent information on CKM parameters.

\subsection{$B_s \to K^*\pi$}

The isospin decomposition of the two $\Delta S=0$ $B_s\to K^* \pi $ decay
amplitudes is:
\begin{align}
\begin{split}
A_s(K^{*+}\pi^-)&=A^s_{3/2}-\sqrt{2}B^s_{1/2}~,\\
A_s(K^{*0}\pi^0)&=\sqrt{2}A^s_{3/2} + B^s_{1/2}~,
\end{split}
\end{align}
where the superscript $s$ denotes $B_s$ instead of $B^0$ and subscripts denote the
isospin of both
the transition operator and the final $K^*\pi$ state.
Since in $\Delta S=0$ decays the QCD penguin operator behaves as $\Delta I=1/2$ it is
contained only in $B^s_{1/2}$. On the other hand, the amplitude
\beq\label{As3/2}
3A^s_{3/2} = A_s(K^{*+}\pi^-) + \sqrt{2}A_s(K^{*0}\pi^0)
\eeq
is pure tree when neglecting EWP contributions, thus providing information on
$\gamma$. Defining a ratio of $\bar B_s$ and $B_s$ amplitudes
(denoted $R_d$ in \cite{Ciuchini:2006st}),
\beq\label{Rs3/2}
R^s_{3/2} \equiv \frac{\bar A^s_{3/2}}{A^s_{3/2}}~,
\eeq
one now has
\beq\label{argRs3/2}
\Phi^s_{3/2}\equiv -\frac{1}{2}{\rm arg}(R^s_{3/2}) = \gamma~.
\eeq

\subsection{$B_s\to K^*\bar K$ and $B_s\to \bar K^* K$}
These $\Delta S=1$ decays involve two independent pairs of isospin amplitudes,
\begin{align}
\begin{split}
A_s(K^{*+}K^-) &= A^s_1 + B^s_0~,\\
A_s(K^{*0}\bar K^0) &= A^s_1 - B^s_0~,
\end{split}
\end{align}
and
\begin{align}
\begin{split}
A_s(K^{*-}K^+) &=A'^s_1 + B'^s_0~,\\
A_s(\bar K^{*0} K^0) &= A'^s_1 - B'^s_0~.
\end{split}
\end{align}
Thus, one has
\begin{align}
\begin{split}
2A^s_1 &= A_s(K^{*+}K^-) + A_s(K^{*0}\bar K^0)~,\\
2A'^s_1 &= A_s(K^{*-}K^+) + A_s(\bar K^{*0} K^0)~.
\end{split}\label{As1}
\end{align}
Defining for each of these processes a ratio of $\bar B_s$ and
$B_s$ amplitudes,
\beq\label{Rs1}
R^s_1 \equiv \frac{\bar A^s_1}{A^s_1}~,~~~~~R'^s_1 \equiv
\frac{\bar A'^s_1}{A'^s_1}~,
\eeq
one obtains two new independent equations for $\gamma$,
\beq\label{argRs1}
\begin{split}
\Phi^s_1 &\equiv -\frac{1}{2}{\rm arg}(R^s_1) = \gamma~,\\
\Phi'^s_1 &\equiv -\frac{1}{2}{\rm arg}(R'^s_1) = \gamma~.
\end{split}
\eeq

\section{CKM constraints including electroweak penguin amplitudes}\label{sec:ewp}
In section \ref{sec:decomposition} we have neglected $\Delta S=1, \Delta I=1$
and $\Delta S=0, \Delta I=3/2$ EWP contributions. In this limit measurements
of the ratios $R^{(s)}_I$ in (\ref{argRI}), (\ref{argRs3/2}) and (\ref{argRs1})
determine $\gamma$.
The inclusion of EWP operators modifies these relations
since these operators involve different weak phase than the tree operators.
These effects are important in the $\Delta S=1$ relations (\ref{argRI}) and (\ref{argRs1}),
where EWP contributions  are CKM-enhanced, and are negligible in the $\Delta S=0$
relation (\ref{argRs3/2}). We will first obtain a general constraint in the
$(\bar\rho,\bar\eta)$ plane~\cite{Eidelman:2004wy}
following from fixed values of $\Phi^{(s)}_I\equiv -\frac{1}{2}{\rm arg}(R^{(s)}_I)$.

Let us study the effect of EWP operators on obtaining CKM constraints in
$\Delta S=1$ decays.
The dominant $(V-A)$ EWP operators, $O^s_9$ and $O^s_{10}$, in the $\Delta S=1$
effective Hamiltonian~\cite{Buchalla:1995vs} are related to current-current operators,
$O^s_1\equiv[\bar s b]_{V-A}[\bar uu]_{V-A}$
and $O^s_2\equiv [\bar ub]_{V-A}[\bar s u]_{V-A}$, through operator relations,
\beq
O^s_{9,10}=\frac{3}{2} O^s_{1,2} +[{\rm operators~with~}~\Delta I=0]~.
\eeq
Neglecting EWP operators, $O^s_7$ and $O^s_8$, involving small Wilson coefficients,
the $\Delta I=1$ part of the $\Delta S=1$ weak Hamiltonian can be rewritten as
\beq\label{H}
\begin{split}
H^s_{\Delta I=1}&=\left(\lambda_u^sC_+ -\frac{3}{2}\lambda_t^s
C_+^{\rm{EWP}}\right)O^{\Delta I=1}_+\\
&+\left(\lambda_u^s C_- - \frac{3}{2}\lambda_t^sC_-^{\rm{EWP}}\right)
O^{\Delta I=1}_-,
\end{split}
\eeq
where $\lambda^s_{u(t)}\equiv V^*_{u(t)b}V_{u(t)s}$, $O^{\Delta I= 1}_\pm \equiv
\frac12 (O^s_1 \pm O^s_2)$, and {$C_\pm\equiv C_1\pm C_2$,
$C_\pm^{\rm{EWP}}\equiv C_9\pm C_{10}$ are sums and differences of Wilson coefficients.

Terms in (\ref{H}) involving $C_\pm^{\rm{EWP}}$ introduce in $\Delta I=1$ amplitudes
a weak phase different from $\gamma$, with coefficients depending on hadronic
matrix elements for $O^{\Delta I=1}_-$ and $O^{\Delta I=1}_+$.
Using a relation between Wilson coefficients which holds up to $1\%$
corrections~\cite{Buchalla:1995vs},
\beq
\frac{C_+^{\rm{EWP}}}{C_+}=-\frac{C_-^{\rm{EWP}}}{C_-}~,\label{ratios}
\eeq
one obtains a generic expression for the four ratios $R_{1/2}, R_{3/2}, R^s_1$ and
$R'^s_1$ in Eqs.~(\ref{RI}) and (\ref{Rs1}),
\beq\label{R}
R^{(s)}_I=e^{-2i[\gamma+\arg(1+\kappa)]}\frac{1+c_\kappa^* r^{(s)}_I}
{1+c_\kappa r^{(s)}_I}~.
\eeq
Here we define
\beq\label{kappa-def}
c_\kappa\equiv \frac{1-\kappa}{1+\kappa}~, \qquad
\kappa\equiv -\frac{3}{2}\frac{C_+^{\rm{EWP}}}{C_+}\frac{\lambda_t^s}{\lambda_u^s}~,
\eeq
\begin{eqnarray}\label{r}
r^{(s)}_I \equiv \frac{\langle f_I |C_-O^{\Delta I=1}_-|B_{(s)}\rangle}{\langle
f_I|C_+O^{\Delta I=1}_+|B_{(s)}\rangle}~.
\end{eqnarray}

The parameter $\kappa$ depends only on calculable Wilson
coefficients and on CKM parameters. In order to illustrate the
sizable shift in $\Phi^{(s)}_I\equiv -\frac{1}{2}{\rm
arg}(R^{(s)}_I)$ caused by this parameter alone, we use the central
values for CKM parameters~\cite{CKMfitter} and next to leading order
(NLO) values for Wilson coefficients at $\mu=m_b=4.8$ GeV,
$C_1(m_b)=-0.178$, $C_2(m_b)=1.079$, $C_9(m_b)=-0.0102$,
$C_{10}=0.0017$. We find \beq\label{kappa}
\begin{split}
\kappa&=\frac{\lambda_t^s}{\lambda_u^s} (1.404\pm0.038)\times 10^{-2}\\
&=-0.35+0.56 i~,
\end{split}
\eeq
where the error in the brackets corresponds to varying the scale $\mu$ in the
NLO Wilson coefficients in the range $m_b/2 \le \mu \le m_b$.
In the second line we give the
result for central values of Wilson coefficients and CKM elements. This
value of $\kappa$ translates into $\arg(1+\kappa)=41^\circ$.

A nonzero value of the parameter $r^{(s)}_I$ leads to an additional
shift in $\Phi^{(s)}_I$, given by
$-\frac{1}{2}{\rm arg}[(1+c_\kappa^* r^{(s)}_I)/(1+c_\kappa r^{(s)}_I]$.
A given value of the observable $\Phi^{(s)}_I$ can be shown to
imply the following constraint in the $(\bar\rho,\bar\eta)$ plane
(we use $\lambda=0.227$):
\beq\label{rho-eta-general}
\frac{\bar\eta +(\bar\rho +C)t}{(\bar\rho + C) -\bar\eta t}
= \tan\Phi^{(s)}_I~.
\eeq
Here we define
\beq
C\equiv\frac{3}{2}\frac{C_+^{\rm EWP}}{C_+}\frac{1-\lambda^2/2}
{\lambda^2} = -0.27~,
\eeq
\beq
t \equiv \frac{1 + t_+ t_- - \sqrt{(1+t_+^2)(1+t_-^2)}}{t_+ - t_-}~,
\eeq
\beq\label{t+-}
\begin{split}
&t_{\pm} \equiv \\
&\frac{(\bar\rho^2+\bar\eta^2-C^2){\rm Im}(r^{(s)}_I) \mp
2 C\bar\eta {\rm Re}(r^{(s)}_I)}
{(\bar\rho+C)^2 + \bar\eta^2 + (\bar\rho^2+\bar\eta^2-C^2){\rm Re}(r^{(s)}_I) \pm
2 C\bar\eta{\rm Im}(r^{(s)}_I)}~.
\end{split}
\eeq

For a small value of $r^{(s)}_I$ and for a given value of
the observable $\Phi^{(s)}_I$, one obtains the
following constraint,
\beq\label{rho-eta}
\bar\eta = \tan\Phi^{(s)}_I
\left[\bar\rho + C\left(1 - 2{\rm Re}(r^{(s)}_I)\right) + {\cal O}(r_I^{(s)2})\right]~.
\eeq
This describes a straight line in the $(\bar\rho,\bar\eta)$ plane
(cf. Fig. \ref{fig3/2}),
with a slope $\tan\Phi^{(s)}_I$ and an intercept
$\bar\rho_0=-C[1-2{\rm Re}(r^{(s)}_I)]$ at $\bar\eta=0$.
A theoretical error $\delta r_I^{(s)}$ in $r_I^{(s)}$ translates into an uncertainty
of $\pm 2|C|\delta r_I^{(s)}$ in the intercept $\bar\rho_0$ but no uncertainty in the
slope $\tan\Phi^{(s)}_I$ which is measured through the ratio $R_I^{(s)}$.
Assuming for illustration a negligible value of $r_I^{(s)}$, one may estimate the
slope required in the Standard Model by choosing central values of $(\bar\rho,\bar\eta)$
from a CKM fit~\cite{CKMfitter},
$\bar\rho=0.20,~\bar\eta=0.34$.
This implies a slope
$\tan\Phi^{(s)}_I=-5.0$,
which is quite sensitive to the value of $r_I^{(s)}$.

\begin{figure}
\includegraphics[width=8.7cm]{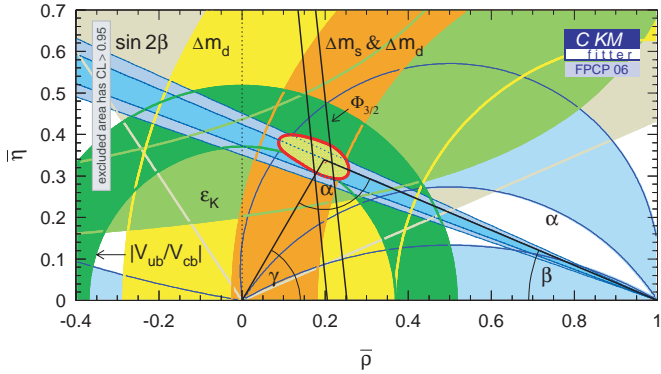}
\caption{$1\sigma$ constraint  in the $\bar \rho-\bar \eta$ plane (two almost vertical
parallel black lines) from a precisely measured $\Phi_{3/2}$ in $B\to K\pi\pi$, taking
values for $r_{3/2}$ as in \eqref{r_{3/2}}.
All other constraints are taken from
\cite{CKMfitter}.}\label{fig3/2}
\end{figure}

A similar treatment of EWP contributions can be applied to $\Delta S=0$
processes involving CKM factors $\lambda^d_{u(t)}$. In the isospin symmetry
 limit the $\Delta S=0$ part of
$O_-$ is pure $\Delta I=1/2$, hence $O^{\Delta I=3/2}_-=0$.
Consequently, in $B_s\to (K^*\pi)_{I=3/2}$ one has
$r^s_{3/2}=0$.~\footnote[1]{This observation has been
overlooked in Ref.~\cite{Ciuchini:2006st}.} This implies that there
is no hadronic uncertainty in the CKM constraint from the ratio
$R^s_{3/2}$ in $B_s\to K^*\pi$, aside from tiny corrections from the
operators $O_7$ and $O_8$ which we have neglected.
The parameter $\kappa'$ in
$\Delta S=0$ decays, whose complex phase is related to $\gamma$, is
of order a few percent [see Eqs.~(\ref{kappa-def}) and (\ref{kappa})],
\beq\label{kappa'}
|\kappa'|=-\frac{3C^{\rm EWP}_+}{2C_+}\frac{|\lambda^d_t|}{|\lambda^d_u|}
= (1.40 \pm 0.04)\times 10^{-2}\frac{\sin\gamma}{\sin\beta}~.
\eeq
Thus, the dependence of the shift ${\rm arg}(1+\kappa')$ on $\gamma$ is very
small and calculable in terms of $\gamma$.

Another case where $r=0$ holds in a symmetry limit is  $B\to (K\pi)_{I=3/2}$,
where the $K$ and $\pi$ mesons are in an S-wave and must be in a symmetric
SU(3) state~\cite{Neubert:1998pt,Gronau:1998fn}. This SU(3) argument does not hold
in $B\to (K^*\pi)_{I=1/2,3/2}$~\cite{Gronau:2003yf}, nor does it hold in
$B_s\to (K^*\bar K)_{I=1}$ and $B_s\to (\bar K^* K)_{I=1}$.
We will study now the values of $r^{(s)}_I$ for these decays within flavor SU(3).
Theoretical errors in these values lead to uncertainties in the resulting CKM
constraints.

\subsection{Ratios $r_I$ and CKM constraints in $B\to K^*\pi$}

The two ratios $R_I$ in Eq.~(\ref{RI}), providing independent pieces of
information on $\gamma$, are given by Eq.~(\ref{R}) with $r=r_I$ given by
\beq\label{rI}
r_I \equiv \frac{\langle (K^*\pi)_I |C_-O^{\Delta I=1}_-|B\rangle}{\langle
(K^*\pi)_I|C_+O^{\Delta I=1}_+|B\rangle}~,~~~~~I=1/2, 3/2~.
\eeq
The ratio $r_{3/2}$ was estimated in~\cite{Ciuchini:2006kv}, based on
factorization and assuming certain input values for $B$-to-light-mesons form
factors. Here we wish to present a different approach
based on flavor SU(3) for calculating both $r_{3/2}$ and $r_{1/2}$.
Using flavor SU(3) $r_I$ may be calculated from tree-dominated
strangeness-conserving $B$ decays, which are CKM-enhanced relative to
tree amplitudes in $B\to K^*\pi$, and which have already been measured.
Furthermore, one may apply Eqs.~(\ref{R}) and (\ref{rI}) to
$|(K^*\pi)_X\rangle$, an arbitrary superposition of
$I=1/2$ and $I=3/2$ $K^*\pi$ states. The corresponding
ratio of hadronic matrix elements will be denoted $r_X$.
One is searching for a linear superposition of isospin states
which leads to a small value of $r_X$ in order to obtain a small
uncertainty in CKM parameters.

The operators $O^{\Delta I=1}_+$ and $O^{\Delta I=1}_-$ transform as
$\mathbf{\overline{15}}$ and $\mathbf{6}$ representations of SU(3),
respectively, while a general $K^*\pi$ state is a combination of $\mathbf{8_S},
\mathbf{8_A}, \mathbf{10}, \mathbf{\overline{10}}$ and
$\mathbf{27}$~\cite{Gronau:2000az,Paz:2002ev}.
Thus, the numerator in $r_X$ involves in general a linear combination of three
reduced matrix elements, $\langle \mathbf{8_S} |\mathbf{6}|\mathbf{3}\rangle,
\langle \mathbf{8_A} |\mathbf{6}|\mathbf{3}\rangle, \langle \mathbf{10}
|\mathbf{6}|\mathbf{3}\rangle$,
and the denominator involves a combination of four matrix elements,
$\langle \mathbf{8_S} |\mathbf{\overline{15}}|\mathbf{3}\rangle,
\langle \mathbf{8_A} |\mathbf{\overline{15}}|\mathbf{3}\rangle,
\langle \mathbf{\overline{10}} |\mathbf{\overline{15}}|\mathbf{3}\rangle,
\langle \mathbf{27} |\mathbf{\overline{15}}|\mathbf{3}\rangle$.
The same reduced matrix elements occur in $\Delta S=0$ amplitudes.
One is seeking two sums of  $\Delta S=0$ amplitudes which are given
by the same two combinations in the numerator and denominator of $r_X$.

\begin{table*}
\newcommand{\aerr}[4]   {\mbox{${{#1}^{+ #2}_{- #3}\pm #4}$}}
\newcommand{\berr}[4]   {\mbox{${{#1}\pm #2^{+ #3}_{- #4}}$}}
\newcommand{\cerr}[3]   {\mbox{${{#1}^{+ #2}_{- #3}}$}}
\newcommand{\aerrsy}[5] {\mbox{${{#1}^{+ #2 + #4}_{- #3 - #5}}$}}
\newcommand{\aerrsyt}[7] {\mbox{${{#1}^{+ #2 + #4 + #6}_{- #3 - #5 - #7}}$}}
\newcommand{\berrsyt}[6] {\mbox{${{#1}\pm #2^{+ #3 + #5}_{- #4 - #6}}$}}
\newcommand{\err}[3]   {\mbox{${{#1}\pm{#2}\pm{#3}}$}}
\begin{minipage}{0.50\textwidth}
\caption{\footnotesize{Branching ratios for $B\to \rho\pi$ and $B\to K^*K$ decays,
in units of $10^{-6}$, taken from Ref.~\cite{HFAG} unless quoted otherwise.}}
\label{table-1}
\begin{ruledtabular}
\begin{tabular}{llll}
& Mode & Branching Ratio \\
\hline
$B^+\to$
&$\rho^0\pi^+$ & $\cerr{8.7}{1.0}{1.1}$ \\
&$\rho^+\pi^0$ & $\cerr{10.8}{1.4}{1.5}$ \\
&$\overline{K^{*0}}K^+$ & $<5.3$ \\
\hline
$B^0 \to$
& $\rho^{\pm} \pi^\mp$ & $24.0 \pm 2.5$ \\
& $\rho^+\pi^-$ & $16.0\pm 2.0$
\footnote{We take an average of $19.5 \pm 5.0$~\cite{Wang:2004va}
and $15.3 \pm 2.2$~\cite{Aubert:2006fg}.} \\
& $\rho^-\pi^+$ & $7.6 \pm 1.3$
\footnote{We take an average of $9.6\pm 3.4$~\cite{Wang:2004va}
and $7.3\pm 1.4$~\cite{Aubert:2006fg}.} \\
&$\rho^0\pi^0$ & $\cerr{1.83}{0.56}{0.55}$ \\
&$K^{*0}\overline{K^0}$& $<1.9$ \\
\end{tabular}
\end{ruledtabular}
\end{minipage}
\end{table*}

The case of $r_{3/2}$ is particularly simple since the state
$(K^*\pi)_{I=3/2}$ contains only two pieces transforming as
$\mathbf{10}$ and $\mathbf{27}$ of SU(3). The transformation
properties of the operators imply that only the $\mathbf{10}$ and
$\mathbf{27}$ pieces contribute to the numerator and denominator,
respectively. Thus, $r_{3/2}$ is proportional to a ratio of
corresponding reduced matrix elements, $\langle \mathbf{10}
|\mathbf{6}|\mathbf{3}\rangle/ \langle \mathbf{27}
|\mathbf{\overline{15}}|\mathbf{3}\rangle$. The numerical
coefficient multiplying this ratio can be read off SU(3)
Clebsch-Gordan tables in~\cite{Paz:2002ev} (after translating into
our phase convention). These tables can also be used to express
$\langle \mathbf{10} |\mathbf{6}|\mathbf{3}\rangle$ and $\langle
\mathbf{27} |\mathbf{\overline{15}}|\mathbf{3}\rangle$ in terms of
$\Delta S=0$ amplitudes, \beq\label{r3/2all}
\begin{split}
&r_{3/2}  =  \\
&\frac{[A(\rho^+\pi^0) - A(\rho^0\pi^+)]
-\sqrt2 [A(K^{*+}\bar K^0) -A(\bar K^{*0} K^+)]}{A(\rho^+\pi^0) + A(\rho^0\pi^+)}~.
\end{split}
\eeq

This expression can be simplified by neglecting the $\Delta S=0$ QCD penguin amplitude
given by the second term in the numerator, and by assuming that the strong phase difference
between the two amplitudes in the remaining term is small, as this phase is expected to be
suppressed by $1/m_b$ and $\alpha_s(m_b)$~\cite{Beneke:1999br,Beneke:2003zv,Bauer:2004tj}.
This is supported by studies of QCD penguin amplitudes (including charming penguins)
in $B\to \rho\pi$ which have been found to be small, with a penguin-to-tree ratio of
about 0.2~\cite{Gronau:2004tm}. Signs of
color-allowed amplitudes are assumed to be given by factorization.
Using branching ratios given in Table I, one finds
\beq
\begin{split}
r_{3/2} &=\frac{|\sqrt{{\cal B}(\rho^+\pi^0)} - \sqrt{{\cal B}(\rho^0\pi^+)}|}
{\sqrt{{\cal B}(\rho^+\pi^0)} + \sqrt{{\cal B}(\rho^0\pi^+)}} \\
&= 0.054 \pm 0.045 \pm 0.023~.
\label{r_{3/2}}
\end{split}
\eeq
The first error is caused by experimental errors in $B\to\rho\pi$ branching ratios.
The second error, due to SU(3) breaking, is calculated by allowing an uncertainty
of $30\%$ in each of the reduced matrix elements entering the physical amplitudes.

The value (\ref{r_{3/2}}), obtained by applying SU(3) to $B\to\rho\pi$ branching ratios,
may be compared with an estimate based on naive factorization~\cite{Ciuchini:2006kv}
in which we include a color factor,
\beq
\begin{split}
r_{3/2} &=\frac{C_-}{C_+}\frac{(1-1/N_c)}{(1+1/N_c)}\frac
{(f_\pi A_0^{BK^*}-f_{K^*}F_0^{B\pi})}{(f_\pi A_0^{BK^*}+f_{K^*}F_0^{B\pi})}\\
&=0.012 \pm 0.083~.
\end{split}
\eeq
We used the following values for decay constants and form factors~\cite{Beneke:2003zv},
$f_\pi=131$~MeV,~$f_{K^*}=218\pm 4$~MeV,~$F_0^{B\pi}=0.28\pm 0.05$,
$A_0^{BK^*}=0.45\pm 0.07$. Note that naive factorization may be a reasonable approximation
because the ratio $r_{3/2}$ defined in \eqref{rI} does not involve QCD penguin contributions.

A bound on the error in $r_{3/2}$ caused by neglecting a difference
of two $\Delta S=0$ QCD penguin amplitudes in (\ref{r3/2all}) can be
obtained in terms of upper bounds on branching fractions for $B^+\to
K^{*+}\bar K^0$ and $B^+\to \bar K^{*0}K^+$. Although current upper
bounds are not very useful (see see Table I), we expect the bounds
to improve in the future, such that the error caused by neglecting
these terms will be at most at the level of SU(3) breaking.

The error in $r_{3/2}$ affects the CKM constraint (\ref{rho-eta}) through
the term involving ${\rm Re}(r_{3/2})$. The error from neglecting a strong
phase difference between $A_{\rho^+\pi^0}$ and $A_{\rho^0\pi^+}$ is expected
to be very small, since ${\rm Re}(r_{3/2})$ depends quadratically
on this phase. This is gratifying since the size of $1/m_b$ suppressed strong
phases cannot be reliably calculated~\cite{Beneke:1999br,Cheng:2004ru}.
Information on the above phase is provided by the isospin pentagon
relation~\cite{Nir:1991cu},
\beq
\begin{split}
&A(\rho^+\pi^0)+ A(\rho^0\pi^+)=\\
& \frac{1}{\sqrt{2}}\big(A(\rho^+\pi^-)+
A(\rho^-\pi^+)\big)+\sqrt{2}A(\rho^0\pi^0)~.
\label{A_0}
\end{split}
\eeq
Relative phases between amplitudes on the right-hand-side can be measured through a Dalitz
plot analysis of $B^0\to \pi^+\pi^-\pi^0$~\cite{Snyder:1993mx,Aubert:2004iu}.
Assuming that phases between $B\to\rho\pi$ amplitudes can be neglected, and using branching
ratios from Table I and a lifetime ratio~\cite{HFAG}, $\tau_+/\tau_0 = 1.076 \pm 0.008$,
Eq.~(\ref{A_0}) reads $6.01 \pm 0.27 = 6.69 \pm 0.38$. This agreement shows that
relative phases between $B\to\rho\pi$ amplitudes are not large.
Assuming in contrast a negative sign for the color-suppressed amplitude $A(\rho^0\pi^0)$,
for which factorization does not hold, would imply $6.01 \pm 0.27 = 2.86 \pm 0.38$ which is
badly violated.

The value of $r_{3/2}$ in (\ref{r_{3/2}}) can now be substituted in Eq.~(\ref{rho-eta}).
The resulting linear constraint in $\bar\rho-\bar\eta$ plane is shown in Fig. \ref{fig3/2},
assuming a precisely measured value for $\Phi_{3/2}$.
The current error in $r_{3/2}$ translates into a very small error of in the
intercept where $\bar\eta=0$,
$\bar\rho_0 = 0.24 \pm 0.03$, but no theoretical
error in the slope which is given by a value measured for 
$\tan\Phi_{3/2}$.
The small error in the intercept, partly from SU(3) breaking in $r_{3/2}$, is linear
in the uncertainty in $r_{3/2}$, and may be reduced only slightly by measuring more
precisely $B\to \rho\pi$ branching ratios.

The calculation of $r_{1/2}$ proceeds in a similar manner to the calculation
of $r_{3/2}$, leading to a larger value of order one. Instead, one may search
for a superposition of $I=1/2$ and $I=3/2$ $K^*\pi$ states for which $r_X$
is small. Using
\beq\label{I_X}
|(K^*\pi)_X\rangle = \frac{1}{\sqrt{5}}(|I=1/2\rangle - 2|I=3/2\rangle)~,
\eeq
we find
\beq
\begin{split}
& r_X = \\
& \frac{A(\rho^+\pi^0) -2 A(\rho^0\pi^+)+\sqrt2 A(\rho^0\pi^0)+N_X(K^*K)}
{A(\rho^+\pi^0) + \sqrt2 A(\rho^-\pi^+) +\sqrt2 A(\rho^0\pi^0)+D_X(K^*K)}~.
\end{split}
\eeq
$\Delta S=0$ QCD penguin and annihilation amplitudes in the numerator and
denominator,
\beq
\begin{split}
N_X(K^*K)\equiv & \big[2A(\bar K^{*0}K^+)-3A(K^{*+}\bar K^0) \\
& - A(K^{*0}\bar K^0)+A(K^{*+}K^-)\big]/\sqrt2~, \\
D_X(K^*K)\equiv& \big[A(K^{*0}\bar K^0)-A(K^{*+}\bar K^0) \\
& +A(K^{*+}K^-)\big]/\sqrt2~,
\end{split}
\eeq
are expected to be no larger than SU(3) breaking corrections
and will be neglected. Assuming small strong phase
differences between $B\to\rho\pi$ amplitudes, and using measured branching
ratios in Table I, we find,
\beq\label{r_X}
r_X = -0.068 \pm 0.057 \pm 0.044~.
\eeq
The first error originates in experimental errors in $B\to\rho\pi$ branching ratios,
while the second error is calculated assuming $30\%$ SU(3) breaking in reduced matrix
elements. The central value of $r_X$ and its error are comparable to $r_{3/2}$. As in the
latter case, this translates to a very small error in the intercept, but no error in
the slope of the linear relation (\ref{rho-eta}) between $\bar\rho$ and $\bar\eta$
provided by a measurement of $\Phi_X$. We define $R_X$ and $\Phi_X$ as in
Eqs.~(\ref{RI}) and (\ref{argRI}) using the $K^*\pi$ state defined in (\ref{I_X}),
\beq
\begin{split}
R_X \equiv & \frac{\bar A_{1/2}-2\bar A_{3/2}}{A_{1/2} - 2A_{3/2}}~~,\\
\Phi_X \equiv  & -\frac{1}{2}{\rm arg}(R_X)~~.
\end{split}
\eeq

The result (\ref{r_X}) may be compared with an estimate based on
naive factorization,
\beq
\begin{split}
r_X &=\frac{C_-}{C_+}\frac{(1-1/N_c)}{(1+1/N_c)}\frac{(2 f_\pi A_0^{BK^*}-
f_{K^*}F_0^{B\pi})}{(2 f_\pi A_0^{BK^*}+f_{K^*}F_0^{B\pi})}\\
&= -0.22 \pm 0.07~,
\end{split}
\eeq
where the error reflects only errors on the assumed form factors.
Our result (\ref{r_X}) using flavor SU(3) agrees within uncertainties with this more
crude approximation which gives a somewhat larger central value.

\subsection{Ratio $r^s_{3/2}$ and determining $\gamma$ in $B_s\to K^*\pi$}

As we have shown using isospin symmetry alone, the parameter $r$ in
$B_s\to K^*\pi$ vanishes, $r^s_{3/2}=0$, because the $\Delta S=0$
part of $O_-$ is pure $\Delta I=1/2$.
The small parameter $\kappa'$ introduces a small shift ${\rm arg}(1+\kappa')$
in $\Phi^s_{3/2}$ away from $\gamma$. Since the shift is calculable in terms of
$\gamma$ [see Eq.~(\ref{kappa'})], the theoretical error in determining $\gamma$
using these processes is below one degree.

Note that measuring $\gamma$ in these processes, and using $B\to
K\pi\pi$ for constraining the point ($\bar\rho,\bar\eta$) to lie on
a straight line with measured slope and intercept, fixes the apex of
the unitarity triangle as the point where the two straight lines
intersect. Thus, in principle, $B\to K\pi\pi$ and $B_s\to K\pi\pi$
alone determine the shape of the unitarity triangle.

\subsection{Ratios $r^s_1, r'^s_1$ and CKM constraints in
$B_s\to K^*\bar K,\bar K^* K$}

In the presence of EWP contributions the two ratios $R^s_1$ and $R'^s_1$ defined
in (\ref{Rs1}) are given by Eq.~(\ref{R}) with
\beq\label{rs}
\begin{split}
r^s_1 &\equiv \frac{\langle K^*\bar K |C_-O^{\Delta I=1}_-|B_s\rangle}{\langle
K^*\bar K|C_+O^{\Delta I=1}_+|B_s\rangle}~,\\
r'^s_1 &\equiv \frac{\langle \bar K^* K |C_-O^{\Delta I=1}_-|B_s\rangle}{\langle
\bar K^* K|C_+O^{\Delta I=1}_+|B_s\rangle}~.
\end{split}
\eeq
We use SU(3) tables in Ref.~\cite{Paz:2002ev} to express these ratios
in terms of $\Delta S=0$ decay amplitudes for nonstrange $B$ mesons,
\beq
\begin{split}
r^s_1 &=\frac{A(\rho^+\pi^-) + A(\rho^-\pi^+) - \sqrt2 A(\rho^0\pi^+) +N_1(K^*K)}
{A(\rho^+\pi^-) - A(\rho^-\pi^+) + \sqrt2 A(\rho^0\pi^+)+ D_1(K^*K)}~,\\
r'^s_1 &=\frac{A(\rho^-\pi^+) + A(\rho^+\pi^-) - \sqrt2 A(\rho^+\pi^0)+N'_1(K^*K)}
{A(\rho^-\pi^+) - A(\rho^+\pi^-) + \sqrt2 A(\rho^+\pi^0)+D'_1(K^*K)}~.
\end{split}
\eeq
Penguin and annihilation terms in the numerators and denominators,
\beq
\begin{split}
N_1(D_1)&\equiv \pm A(\bar K^{*0}K^+) \mp A(K^{*-}K^+) + A(\bar K^{*0}K^0)~,\\
N'_1(D'_1)&\equiv \pm A(K^{*+}\bar K^0) \mp A(K^{*+}K^-) + A(K^{*0}\bar K^0)~,
\end{split}
\eeq
will be assumed to be smaller than SU(3) breaking corrections.

Disregarding phase differences between $B\to\rho\pi$ amplitudes which
have a very small effect on $r^s_1$ and $r'^s_1$
[as we argued in obtaining (\ref{r_{3/2}})], using measured branching
ratios in Table I,
and estimating errors from SU(3) breaking as explained above, we have
\beq\label{rs1exp}
\begin{split}
r^s_1 &= 0.52 \pm 0.10 \pm 0.18~,\\
r'^s_1 &= 0.70 \pm 0.21 \pm 0.41~.
\end{split}
\eeq
Comparing these values with an estimate
based on naive factorization, we find agreement again,
\beq
r^s_1 = r'^s_1 = -\frac{C_-}{C_+}\frac{(1-1/N_c)}{(1+1/N_c)} = 0.70~.
\eeq

We do not expect the errors in (\ref{rs1exp}) to improve by reducing errors in
${\cal B}(B\to\rho\pi)$, as SU(3) breaking introduces a comparable uncertainty.
The values of $r^s_1$ and $r'^s_1$ can be substituted in
Eqs.~(\ref{rho-eta-general})-(\ref{t+-}) to obtain constraints in the
$(\bar\rho,\bar\eta)$ plane, for measured values of $\Phi^s_1$ and $\Phi'^s_1$.
The larger errors in (\ref{rs1exp}) in comparison with those in (\ref{r_{3/2}})
and (\ref{r_X}) imply larger uncertainties in these constraints than in those following
from $\Phi_{3/2}$ and $\Phi_X$.

\section{Measuring magnitudes and phases for quasi two-body decay amplitudes}\label{sec:measuring}
As shown in the previous two sections, new constraints in the $(\bar\rho,\bar\eta)$ plane
can be obtained within each of the three classes of quasi two-body decay processes,
$B\to K^*\pi$, $B_s\to K^* \pi$, and $B_s\to K^*\bar K, \bar K^* K$ and their
charge-conjugates.
This requires measuring both the magnitudes of the amplitudes in a given class 
and their relative phases. This can be achieved through amplitude analyses of
charmless three-body decays which we discuss now.

A three-body $B$ (or $B_s$) decay amplitude into a final state $f$, which is a function
of two Dalitz variables, $s_{12}, s_{13}$,
is expressed as a sum of several Breit-Wigner resonant contributions and a non-resonant term.
Resonant contributions are given by complex constant amplitudes $A_i$ multiplying
Breit-Wigner functions $f_i^{BW}(s_{12},s_{13})$, while the nonresonant
amplitude $A_{\rm NR}$ may vary in the $s_{12},s_{13}$ plane,
\beq\label{BWsum}
A(s_{12},s_{13})=A_{\rm NR}(s_{12},s_{13}) + \sum_i A_i f_i^{BW}(s_{12},s_{13})~.
\eeq
The corresponding amplitude $\bar A(s_{12},s_{23})$, for three-body
$\bar B$ (or $\bar B_s$) decays into a charge-conjugate state $\bar f$, is
given in terms of an amplitude $\bar A_{\rm NR}$ and a set $\bar A_i$
corresponding to charge-conjugate resonances. In general, one has
$\bar A_{\rm NR}\ne A_{\rm NR}$, $\bar A_i\ne A_i$ as each amplitude
may involve two weak phases and two different strong phases. Direct CP
violation in a particular resonant or non-resonant channel would be
implied by $|\bar A_i|\ne |A_i|$ or $|\bar A_{\rm NR}| \ne |A_{\rm NR}|$.

Fitting the event distribution for three body $B$ (or $B_s$) decays to the squared
amplitude (\ref{BWsum}) permits determining the
magnitudes of $A_i$ and their relative phases. A relative phase between two resonant
amplitudes is directly measurable when the two resonances  overlap in the Dalitz plot.
This relative phase can also be measured when there is no overlap between the two
resonances, but each of the two resonances overlaps with a third resonance.
Alternatively, a phase between two resonance amplitudes can be measured
through their interference with the nonresonant amplitude $A_{\rm NR}$.

We will be interested primarily in relative phases between amplitudes associated 
with $K$ meson resonant states. Charmless three-body decays involving $\pi^+\pi^-$ or 
$K^+K^-$ obtain also contributions from $c\bar c$ resonant states, which involve 
relatively small rates and are expected to lead to sizable CP 
asymmetries~\cite{Eilam:1995nz,Bediaga:1998ma,Fajfer:2002ct}.

\subsection{$B\to K\pi\pi$}

We start this discussion with the decays $B\to K\pi\pi$ which are currently
the most feasible ones among the three classes studied in this paper.
Amplitude analyses of $B\to K\pi\pi$, for both charged and neutral $B$ mesons,
have been performed by the Belle and Babar collaborations. Decays $B^+\to K^+\pi^+\pi^-$
have been studied by both Belle~\cite{Garmash:2004wa} and Babar~\cite{Aubert:2005ce}.
An amplitude analysis was made by Babar~\cite{Aubert:2004bt} for
$B^0\to K^+\pi^-\pi^0$~\cite{Chang:2004um}, and by Belle~\cite{Abe:2005kr} for
$B^0\to K_S\pi^+\pi^-$.
The first two processes are self-tagging whereas the third decay involves final
state which is not flavor specific.
These measurements have already provided some useful information which is relevant to
our proposed study. We note that these studies have averaged over the above processes
and their CP-conjugates. The proposed study requires separate amplitude analyses for
$B$ and $\bar B$ decays.

The process $B^+\to K^+\pi^+\pi^-$ gave information about the magnitudes of
amplitudes for $B^+\to K^{*0}(892)\pi^+$ and $B^+\to K^{*0}_0(1430)\pi^+$
and their relative phase~\cite{Garmash:2004wa,Aubert:2005ce}. The statistical
error in the measured relative phase is at a level of $10^\circ$ which is
encouraging. However, this three-body decay provides no information on
a relative phase between two $B\to K^*(892)\pi$ amplitudes where pairs of $K^*$ 
and $\pi$ have different charges.

The decay $B^0\to K^+\pi^-\pi^0$ is more interesting in our context, since it measures
the magnitudes of $A(B^0\to K^{*+}\pi^-)$ and $A(B^0\to K^{*0}\pi^0)$, for both
$K^*(892)$ and $K^*_0(1430)$, as well as the three relative phases among these
amplitudes. Errors in the measured phases are at a level of
$40^\circ$~\cite{Aubert:2004bt}. It would be useful to understand the origin of this
large error in order to reduce it in future studies of this process, and to perform
these measurements separately for $B^0$ and $\bar B^0$.
A study of $B^0\to K^+\pi^-\pi^0$ permits a measurement of the magnitude of 
$R_{3/2}$ but not its phase. Eq.~(\ref{R}) implies that $|R_{3/2}|-1$ is proportional to
${\rm Im}(r_{3/2})$ and vanishes if $r_{3/2}$ is real. 

The study of $B^0\to K_S\pi^+\pi^-$, which is not flavor specific,
is more challenging. In order to measure the relative phase between
$A(K^{*+}\pi^-)$ and $\bar A(K^{*-}\pi^+)$, as required by
Eqs.~(\ref{RI}) and (\ref{argRI}), these amplitudes must interfere
through $B^0$-$\bar B^0$ mixing leading to a common $K_S\pi^+\pi^-$
state. Observing this interference in $e^+e^-$ collisions at the
$\Upsilon(4S)$ requires a time-dependent measurements using
initially tagged $B^0$ or $\bar B^0$ mesons.
The recent time-integrated analysis by Belle~\cite{Abe:2005kr} assumed
no direct CP asymmetry in $B^0\to K^{*+}\pi^-$, summing over initial
$B^0$ and $\bar B^0$. We note that, in fact, an
untagged amplitude analysis does not have to make this assumption, permitting
separate measurements for the magnitudes of $A(K^{*+}\pi^-)$ and $\bar A(K^{*-}\pi^+)$.
However, measuring the relative phase between these amplitudes requires a
time-dependent measurement.

A fourth process in this class, $B^+\to K_S\pi^+\pi^0$, which has not yet
been measured, determines the magnitudes of
the four amplitudes, $A(K^{*0}\pi^+)$, $A(K^{*+}\pi^0)$, $A(K^{*0}_0\pi^+)$,
$A(K^{*+}_0\pi^0)$, and their relative phases. Finally, a very difficult mode
which is not needed is $B^0\to K_S\pi^0\pi^0$, where measuring the phase
difference between $A(K^{*0}\pi^0)$ and $\bar A(\bar K^{*0}\pi^0)$ would require
time-dependence.

In order to apply Eq.~(\ref{rho-eta}), the linear constraint in the
$(\bar\rho,\bar\eta)$ plane, where $r_{3/2}$ is given in
(\ref{r_{3/2}}),  it is sufficient to perform amplitude analyses for
merely two processes involving neutral $B$ decays, $B^0\to
K^+\pi^-\pi^0$ and $B^0\to K_S\pi^+\pi^-$. Time-dependence in the
second process is crucial. The first process measures the magnitudes
of $A(K^{*+}\pi^-)$ and $A(K^{*0}\pi^0)$, their relative phase, and the
corresponding CP-conjugate quantities, but not the phase difference
between $B^0$ and $\bar B^0$ decays. [Here and below $K^*$ denotes
both $K^*(892)$ and $K^*_0(1430)$]. The second process measures the
magnitude of $A(K^{*+}\pi^-)$ and its CP-conjugate, and the
relative phase between these two amplitudes. This set of
measurements determining the complex ratio $R_{3/2}$ defined in
Eqs.~(\ref{quad}) and (\ref{RI}),  is over-complete since
$|A(K^{*+}\pi^-)|$ and its CP-conjugate are measured both in $B^0\to
K^+\pi^-\pi^0$ and in $B^0\to K_S\pi^+\pi^-$.

Charged $B$ decays, $B^{\pm}\to K^{\pm}\pi^{\pm}\pi^{\mp}$ and $B^{\pm}\to
K_S\pi^{\pm}\pi^0$ provide further constraints on CKM parameters using the measurable
ratio $R_X$ of $B\to K^*\pi$ and $\bar B\to K^*\pi$ decay amplitudes. This, together
with the constraint from $R_{3/2}$, leads to a highly constraining set of measurements
for $\bar\rho$ and $\bar\eta$. Since the four physical $B\to K^*\pi$ amplitudes are not
mutually independent [see the quadrangle relation Eq.~(\ref{quad})],
we propose studying $B\to K^*\pi$ amplitudes in terms of the isospin
amplitudes $B_{1/2}$, $A_{3/2}$ and $A_X$, where $X$ corresponds to the state defined
in (\ref{I_X}). In order to demonstrate the extent to which these CKM constraints are
over-deterministic, thereby permitting a precise constraint on the point
($\bar\rho,\bar\eta$), we now count the number of parameters and observables.

We have a total of eight parameters, the magnitudes of $B_{1/2}$ and
its CP conjugate $\bar B_{1/2}$, the magnitudes of $A_{3/2}$ and
$A_X$, the three relative phases among these four amplitudes, and a
CKM ratio $\bar\eta/(\bar\rho+C)$. [The CP conjugates $\bar A_{3/2}$
and $\bar A_X$ are not independent parameters and are given by
Eqs.~(\ref{RI}),~(\ref{argRI}) and (\ref{rho-eta}).] These eight
parameters can be used to fit seventeen observables consisting of
$|A(K^{*0}\pi^{\pm}|$ obtained from $B^+\to K^+\pi^+\pi^-$,
magnitudes of $A(K^{*0}\pi^\pm)$ and $A(K^{*\pm}\pi^0)$ and their
relative phases obtained from $B^{\pm}\to K_S\pi^{\pm}\pi^0$,
magnitudes of $A(K^{*+}\pi^-), A(K^{*0}\pi^0)$, their CP conjugates
and their relative phases obtained from $B^0\to
K^{\pm}\pi^{\mp}\pi^0$, and magnitudes and relative phase for
$A(B^0\to K^{*+}\pi^-)$ and $A(\bar B^0\to K^{*-}\pi^+)$ obtained
from time dependent  $B^0\to K_S\pi^+\pi^-$. We have not included in
this counting the decay $B^0\to K_S\pi^0\pi^0$ which is most
challenging.

\subsection{$B_s \to K\pi\pi$}

The weak phase $\gamma$ can be determined using Dalitz plot analyses for
$B_s\to K^{\pm}\pi^{\mp}\pi^0$ and $B_s\to K_S\pi^{\pm}\pi^{\mp}$. These studies
permit extracting the magnitudes $A_s(K^{*+}\pi^-), A_s(K^{*0}\pi^0)$, their
CP conjugates and relative phases between these amplitudes. This leads through
Eqs.~(\ref{As3/2})-(\ref{argRs3/2}) to a measurement of  the phase
$\Phi^s_{3/2}$, which gives $\gamma$ with  high theoretical precision, as has been
discussed in Section III.B.

In contrast to the case of $B^0\to K_S\pi^+\pi^-$ produced at the $\Upsilon(4S)$,
the above measurements can be performed with $B_s\to K_S\pi^+\pi^-$ produced at hadron
colliders without the need for flavor tagging and time-dependence.
Because of the lack of quantum coherence between $B_s$ and $\bar B_s$ produced
in pairs, the charge-averaged time-integrated decay rate for decays into a common state
$f\equiv K_S\pi^+\pi^-$ involves an interference term proportional to the width difference
$\Delta\Gamma_s$ in the $B_s$ system, for which one expects
$y_s\equiv \Delta\Gamma_s/2\Gamma_s=0.12\pm 0.05$~\cite{Lenz:2004nx}.

The untagged integrated decay distribution is given by,
\beq
\begin{split}\label{sumq}
&\frac{d^2\Gamma(B_s\to f)}{ds_{12}ds_{13}}~+~\frac{d^2\Gamma(\bar B_s\to f)}{ds_{12}ds_{13}}
=\\
&\frac{1}{\Gamma(1-y_s^2)}
\Big[\big(|A|^2+|\bar A|^2\big)-2 y_s \Re\Big(\frac{q}{p}\bar A A^*\Big)\Big]~,
\end{split}
\eeq
where $A\equiv A(B_s\to f),~\bar A\equiv A(\bar B_s\to f)$, $q/p\simeq 1$.
Assuming that a reasonably accurate measurement for $y_s$ will exist by the
time an amplitude analysis will be performed for this decay, the relative
phase between $A_s(K^{*+}\pi^-)$ and $\bar A_s(K^{*-}\pi^+)$ can
be measured through the interference term involving $y_s$.
Otherwise, a time-dependent measurements of this decay will be required.

In order to show that the above relative phase is measurable using untagged $B_s$,
consider the contributions of $A_s(K^{*+}\pi^-)$ and $\bar A_s(K^{*-}\pi^+)$ to
$A$ and $\bar A$ in (\ref{sumq}). Using the dependence of the Breit-Wigner
functions $f^{BW}_{K^{*+}}$ and $f^{BW}_{K^{*-}}$ on $s_{12}$ and $s_{13}$,
the untagged decay distribution (\ref{sumq}) provides four observables
(the real part of the interference term provides two observables) which determine
the magnitudes of $A_s(K^{*+}\pi^-)$ and $\bar A_s(K^{*-}\pi^+)$ and their relative
phase. While in reality this relative phase may be affected by interference with
other resonant or non-resonant terms in the amplitude, this proves that,
once $y_s$ is given, this phase can be measured through an untagged amplitude
analysis of $B_s\to K_S\pi^+\pi^-$.

\subsection{$B_s\to K\bar K\pi$}

As noted above, the CKM constraints following from amplitude
analyses of $B_s\to K\bar K\pi$ decays are less precise than those
following from studies of $B\to K\pi\pi$ and $B_s\to K\pi\pi$. This
is due to theoretical errors in the hadronic electroweak penguin
parameters, $r^s_1$ and $r'^s_1$ [Eq.~\eqref{rs1exp}], which are larger
than in $r_{3/2}$ [Eq.~\eqref{r_{3/2}}], $r_X$ [Eq.~\eqref{r_X}] and $r^s_{3/2}$
[see discussion in Section III.B].

In order to obtain a CKM constraint related to the phase $\Phi^s_1$, for instance, one must
measure the amplitudes in (\ref{As1}), for $B_s\to K^{*+}K^-$ and $B_s\to K^{*0}\bar K^0$, their
charge-conjugates, and the three relative phases between these amplitudes. This can be
achieved by amplitude analyses for a pair of processes belonging to this class.
For instance, using $B_s\to K^+K^-\pi^0$ one can measure the magnitudes of $A(B_s \to K^{*+}K^-)$,
$A(B_s\to K^{*-}K^+)$, their charge-conjugates and the relative phases between these amplitudes.
A study of $B_s\to K^+K_S\pi^-$ permits measurements of the magnitudes of $A(B_s\to K^{*0}\bar
K^0)$, $A(B_s\to K^{*-}K^+)$, $A(\bar B_s\to K^{*0}\bar K^0)$, $A(\bar B_s\to K^{*-}K^+)$ and
the respective relative phases. This information suffices for fixing $\Phi^s_1$.

Decay distributions in $B_s\to K\bar K\pi$ involve twice as many relevant quasi two-body
amplitudes as in $B\to K\pi\pi$ and $B_s\to K\pi\pi$, because $B_s$ and $\bar B_s$ can decay
to a common non-flavor $K^*\bar K$ state. The large number of amplitudes and relative phases
which must be determined in $B_s\to K\bar K\pi$ requires at least as many observables.
While in principle possible, this seems to pose a serious challenge to applying this 
method to $B_s\to K\bar K\pi$ decays.

\section{Conclusion}

We have studied in great detail a method proposed in Ref.~\cite{Ciuchini:2006kv,Ciuchini:2006st}
for obtaining new constraints on CKM parameters using $B_{(s)}\to (K^*\pi)_{I=3/2}$
amplitudes, extending the method to $B\to (K^*\pi)_{I=1/2}$ and to $B_s\to K^*\bar K~(\bar K^*K)$
amplitudes measured in $B\to K\pi\pi$ and $B_s\to K\bar K\pi$, respectively.
Two judiciously chosen isospin amplitudes in $B\to K^*\pi$ have been shown to be
over-constrained by several $B\to K\pi\pi$ amplitude analyses, providing a precise
linear constraint between the CKM parameters $\bar\rho$ and $\bar\eta$. The slope of the
linear relation is a measurable quantity, while the intercept $\bar\rho_0$ where
$\bar\eta=0$ is a calculable quantity involving a theoretical error of
0.03. A study of $B_s\to K^*\pi$ amplitudes in $B_s\to K\pi\pi$ leads to a very accurate
extraction of the weak phase $\gamma$ with a theoretical uncertainty below one degree.

The resulting theoretical precision in determining CKM parameters 
in $B\to K\pi\pi$ and $B_s\to K\pi\pi$  has been shown to be essentially at the level 
of isospin breaking corrections since the method is based on isospin 
symmetry considerations, while flavor SU(3) has been used to 
estimate uncertainties from a subset of small EWP contributions.
A larger hadronic uncertainty from EWP contributions is found in a CKM constraint 
obtained by studying
$B\to K^*\bar K$ and $B\to \bar K^*K$ amplitudes contributing to $B_s\to K\bar K\pi$.

There is one crucial theoretical difference between applying this method to $\Delta S=1$
$B\to K\pi\pi$ and $B_s\to K\bar K\pi$ and applying it to $\Delta S=0$ $B_s \to K\pi\pi$.
The first two classes of processes are dominated by $\Delta I=0$ QCD penguin amplitudes
which are eliminated in the relevant isospin amplitudes. In the Standard Model this
implies a delicate cancellation between physical amplitudes defining the numerators
and denominators of the $\Delta I=1$ observables $R_I$ and $R^s_1 (R'^s_1)$ on
which the method relies. In contrast, in $B_s\to K\pi\pi$ decays the method relies on
measuring the $\Delta I=3/2$ isospin amplitudes which involves dominant tree contributions.
This would seem like a disadvantage of using $B\to K\pi\pi$ and $B_s\to K\bar K\pi$ relative
to $B_s\to K\pi\pi$ for extracting CKM parameters. However, this cancellation in the
Standard Model turns into an advantage when one is searching for New Physics in $\Delta I=1$
operators.

While applications of the method to $B_s$ decays can be foreseen in future
experiments at hadron colliders, data for $B\to K\pi\pi$ are already available
from $e^+e^-$ collisions at the $\Upsilon(4S)$, and should be analyzed in the
manner proposed here.
Amplitude analyses of a few processes in the class $B\to K\pi\pi$
have already been performed, measuring amplitudes and relative phases for $B\to K^*(892)\pi$
and $B\to K^*_0(1430)\pi$~\cite{Garmash:2004wa,Aubert:2005ce,Aubert:2004bt,Abe:2005kr}.
Since the method is based on $\Delta I=1$ amplitudes, a first important step toward
its full implementation is observing
a violation of $\Delta I=0$ QCD penguin dominance in these quasi two-body decays.

This question has been studied recently~\cite{Gronau:2005ax}. It was shown that in all
cases but one $\Delta I=0$ holds well within current experimental errors.
For instance, $\Delta I=0$ dominance implies
\beq
2{\cal B}(B^0\to K_0^{*0}\pi^0) = {\cal B}(B^0\to K_0^{*+}\pi^-)~,
\eeq
which holds experimentally within large errors, in units of $10^{-6}$~\cite{HFAG},
\beq\label{K*pi-Iso}
51.0 \pm 19.8 = 46.6^{+5.6}_{-6.6}~.
\eeq
The exceptional case where $\Delta I=0$ seems to be violated is
the equality,
\beq
2{\cal B}(B^0\to K^{*0}\pi^0) = {\cal B}(B^0\to K^{*+}\pi^-)~,
\eeq
where current experimental values~\cite{HFAG},
\beq\label{K*pi-Iso2}
3.4 \pm 1.6 = 9.8 \pm 1.1~,
\eeq
show a discrepancy of $3.3\sigma$. One would have to watch
carefully whether this discrepancy holds in future measurements.

This method requires performing amplitude analyses of $B\to K\pi\pi$
separately for $B$ and $\bar B$ , as one must measure the ratio of
$\bar B\to \bar K^*\pi$ and $B\to K^*\pi$ amplitudes. The method does
not require observing a direct CP asymmetry in Dalitz plots for
$B\to K\pi\pi$ or an asymmetry in $B\to K^*\pi$ decay rates. We recall that
no time-dependence is needed in order to observe direct CP violation in the
Dalitz plot of $B^0\to K_S\pi^+\pi^-$ through an asymmetry with respect to
exchanging $\pi^+$ and $\pi^-$~\cite{Burdman:1991vt}. We have stressed the
importance of performing a time-dependent Dalitz plot analysis of $B^0\to
K_S\pi^+\pi^-$, which is required in order to measure separately amplitudes for
$B^0\to K^{*+}\pi^-$ and $\bar B^0\to K^{*-}\pi^+$ and their relative phase.

The method presented here for obtaining a linear relation between $\bar\rho$
and $\bar\eta$ in $B\to K\pi\pi$ may be compared with a study of $\gamma$ in
$B\to DK$~\cite{Gronau:1991dp}. The latter method involves
an extremely small theoretical uncertainty from $D^0$-$\bar D^0$
mixing~\cite{Grossman:2005rp} when studying CP-eigenstates and flavor states
in $D$ decays. Applying this method
to non-CP and non-flavor three body $D$ decays such as $D^0\to K_S\pi^+\pi^-$
introduces a theoretical error in $\gamma$ caused by modeling the three body
decay amplitude in terms of a sum of resonant and non-resonant contributions.
Model-dependence in amplitude analyses for $B\to K\pi\pi$ is expected to be
larger than in $D^0\to K_S\pi^+ \pi^-$ because the former processes involve
lower statistics and higher combinatorial backgrounds. Fortunately, the
uncertainty of modeling $B\to K\pi\pi$ is mainly in non-resonant
amplitudes~\cite{Garmash:2004wa,Aubert:2005ce,Aubert:2004bt,Abe:2005kr},
which spread over the entire phase space, but less in
$K^*\pi$ amplitudes which are used in the proposed study.

While measuring $\gamma$ from an interference of tree amplitudes in
$B\to DK$ is most likely to receive only small corrections from New
Physics \cite{Grossman:2005rp,Khalil:2006zb}, the extraction of a
linear constraint between $\bar\rho$ and $\bar\eta$ in $B\to
K\pi\pi$ may be affected
more significantly by such effects. Thus, values for CKM parameters
obtained in the two methods may differ, indicating short distance
$b\to s \bar q q$ operators beyond the Standard Model. The study of
$B\to K\pi\pi$ is insensitive to new $\Delta I=0$ QCD penguin-like
operators which cancel in the ratios $R_I$, but is affected by new
$\Delta I=1$ operators. Such operators are often referred to in the
literature as anomalous electroweak (or Trojan) penguin
operators~\cite{Grossman:1999av}. The sensitivity to such
contributions is high because in the Standard Model $\Delta I=1$
terms in $B\to K^*\pi$ are suppressed relative to $\Delta I=0$
contributions. Other tests for such $\Delta I=1$ operators have been
proposed in terms of isospin sum rules for
rates~\cite{Gronau:1998ep} and CP asymmetries in $B\to K\pi$~\cite{Atwood:1997iw}.

A somewhat similar situation occurs in $B_s$ decays when comparing
the theoretically precise measurement of $\gamma$ in charmless
$B_s\to K\pi\pi$ discussed here with the potentially accurate
measurement of this phase in $B_s\to D^-_sK^+$~\cite{Aleksan:1991nh}. Both methods require $B_s$-$\bar
B_s$ mixing, but no time dependent measurement is required in
$B_s\to K\pi\pi$ due to additional phase information coming from
Dalitz plot interferences. In $B_s\to K\pi\pi$ the measurement of
$\gamma$ follows from studying $\Delta I=3/2$ $\bar b\to \bar u u
\bar d$ tree amplitudes, while in $B_s\to D^-_sK^+$ the phase occurs
in the interference of $\Delta I=1/2$ $\bar b \to \bar c u\bar s$
and $\bar b\to \bar u c \bar s$ tree amplitudes.
Whereas New Physics operators in the latter case are possible in
principle, their effects on the determination of $\gamma$ are less
common and are expected to be much smaller 
than the effects of
potentially new $\Delta I=3/2$ operators contributing in $B_s\to
K\pi\pi$. Such $\Delta S=0$ operators are usually expected in the
same class of models where anomalous $\Delta S=1$ electroweak
penguin operators occur.

\bigskip
\centerline{ACKNOWLEDGMENTS}

M.G. is grateful to the SLAC Theory Group for its kind hospitality.
We wish to thank Denis Dujmic and Roy Briere for helpful discussions.
This work was supported in part by the United States Department of
Energy through Grants No. DOE-ER-40682-143, DEAC02-6CH03000 and DE-AC02-98CH10886,
by the Israel Science Foundation under Grant  No. 1052/04 and by the
German-Israeli Foundation under Grant No. I-781-55.14/2003.


\end{document}